\documentclass[epj]{svjour}
\usepackage{graphics}
\usepackage{xcolor}
\begin{document}
\title{Exploring Dense and Cold QCD in Magnetic Fields}
\author{E. J. Ferrer \and V. de la Incera}
%
%
\institute{University of Texas at El Paso, 500 W. University Ave, El Paso, TX 79968, USA}
\date{Received: date / Revised version: date}
%
\abstract{
Strong magnetic fields are commonly generated in off-central relativistic heavy-ion collisions in the Relativistic Heavy Ion Collider (RHIC) at Brookhaven National Lab and in the Large Hadron Collider at CERN and have been used to probe the topological configurations of the QCD vacua. A strong magnetic field can affect the character and location of the QCD critical point, influence the QCD phases, and lead to anomalous transport of charge. To take advantage of the magnetic field as a probe of QCD at higher baryon densities, we are going to need experiments capable to scan the lower energy region. In this context, the nuclotron-based ion collider facility (NICA) at JINR offers a unique opportunity to explore such a region and  complement alternative programs at RHIC and other facilities. In this paper we discuss some relevant problems of the interplay between QCD and magnetic fields and the important role the experiments at NICA can play in tackling them.
} 
%
\maketitle
\section{Introduction}
\label{intro}

Earthly exploration of quark matter under extreme conditions has been made possible thanks to the heavy-ion collisions experiments carried out by the Relativistic Heavy Ion Collider (RHIC) at Brookhaven National Lab (BNL) and by the Large Hadron Collider (LHC) at CERN. The top center-of-mass energies per nucleon pair reached have been $\sqrt{s}=200$ GeV for the Au-Au collisions at RHIC, and $\sqrt{s}=2.76$ TeV for the Pb-Pb collisions at LHC. These energies are large enough to deconfine the quarks from inside the hadrons and produce the quark-gluon plasma (QGP). The results of these experiments and those planned at LHC,  that will reach $\sqrt{s}=5.5$ TeV, will help to increase the precision of the physical findings for quark-gluon matter in the region of high temperature and low density. On the other hand, the Nuclotron-based Ion Collider Facility (NICA) at JINR will open the possibility to improve the picture of the QCD phase map by extending it to the region of intermediate-to-large densities at lower temperatures.

\begin{figure}
\resizebox{0.25\textheight}{!}{
  \includegraphics{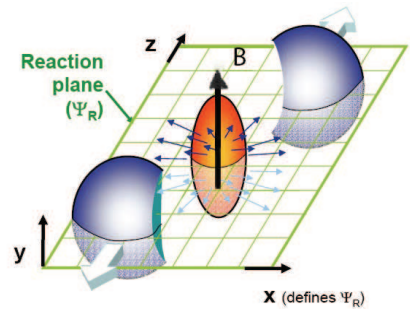}
}
\caption{Generation of magnetic field in off-central heavy-ion collisions. This figure has been modified from Ref. \cite{PPNP75}.}
\label{fig:1}       
\end{figure}
Of particular interest for the understanding of the different phases that can be produced in the above-mentioned experiments are the effects related to the presence of strong magnetic fields. Off-central heavy-ion collision can generate large magnetic fields (see Fig.1).  According to several numerical simulations, off-central Au-Au collisions at RHIC can lead to field strengths of $10^{18}-10^{19}$ G, while the field can be as large as $10^{20}$ G for the off-central Pb-Pb collisions at LHC \cite{NPA803}-\cite{B-simulations-4}. As argued in \cite{NPA803}, these strong magnetic fields, produced during the first instants after a collision, can create the conditions for observable QCD effects. These effects can be prominent because the magnetic fields generated are of the order of or higher than the QCD scale, $eB>\Lambda^2_{QCD}$.

The paper is organized as follows: In Sections \ref{SI} and \ref{SII}, we argue how NICA will allow us to open a new window of exploration of the QCD phase map that will cover a region of high densities and low temperatures not yet reached in the experiment and only present in nature in the core of neutron stars. Important questions as proving/disproving  the existence and location of the QCD critical point, the character of the phase transitions, whether deconfinement and chiral transitions overlap or not, as well as the effects of a strong magnetic field on the various measurable observables are all briefly mentioned. In Section \ref{SIII} we discuss how the planned experiments can help us to probe the topology of different QCD phases and in particular a P-odd inhomogeneous phase of cold and dense quark matter that has been found within model calculations but could be reached in the interval of baryon densities that NICA is planning to explore. This phase is characterized by various anomalous transport properties and it will be important to connect them to observables that can be tested in the experiment. Section \ref{SIV} is dedicated to our concluding remarks.

\section{Exploring the QCD Critical Point and Phase Transitions at NICA} \label{SI}

To set the basis for present and future heavy-ions collision experiments, many investigations have been carried out to improve our theoretical understanding of the QCD phases in the temperature vs baryon chemical potential plane, and particularly, to extend it to include the challenging region of larger densities and lower temperatures. 

Part of these efforts focused on the nature of the chiral and deconfinement phase transitions, the location of the so-called chiral critical point, and the properties in its vicinity, all of which have attracted lot of interest during the past three decades (for review see \cite{deconf} and references therein).

 A priori, it is not evident that confinement and chiral phase transitions have to occur at the same temperature. Early lattice calculations with large quark masses and/or coarse lattices found that deconfinement and chiral transition happen at the same temperature \cite{PRL50}. However, more recent studies \cite{PRC90-6}-\cite{PRC90-6-1}, have found, using the so-called stout staggered quark action and finer lattices, that the two transitions are not as interconnected as had been previously suggested. 
 
Considering that deconfinement and chiral symmetry restoration occur independent from each other, it has been proposed \cite{NPA796} that with increasing density the quark matter could be in a chirally symmetric but confined phase named quarkyonic matter. It was later understood however that chiral symmetry is actually broken in quarkyonic matter, but in a different way, that is, due to the formation of a quarkyonic chiral spiral \cite{NPA7843}, implying that chiral symmetry restoration would occur at even larger densities than expected and only after passing through an inhomogeneous chiral phase. Other candidates for the ground state of quark matter at intermediate-to-large-density are color superconducting (CS) phases, like the 2SC, and crystalline CS phases \cite{RevCS}.

Even though QCD has been established for quite a long time as the theory of the strong interactions, and main properties as confinement and asymptotic freedom are well understood, the theory is strictly applicable only in the region of weak coupling (either very high T or asymptotically high $\mu$).  For stronger values of the coupling, as it is the case at moderately larger density and lower temperatures, one has to rely on model calculations and use nonperturbative tools. In this sense, it is very important to have model predictions for  such a region that can be then tested experimentally, to validate/invalidate the models and the phases theoretically found.

Along this line, NICA experiments offer a unique opportunity to probe this challenging region of the phase diagram, while complementing alternative programs for systematic studies of heavy-ion collisions in the relevant range of collision energies $2\le E\le 11$ GeV. Experiments of the next generation (BES II at RHIC and NICA) should, however, consider the possibility that qualitatively new features could be found at still lower energies. Experimental data on hadron production properties at SPS (CERN) suggest that the transition to the deconfined phase of nuclear matter can occur within the NICA energy range. It is plausible that the range of energy to be explored and the luminosities that will be produced at NICA will allow to probe the QCD transition region, the character of the associated phase transformation, and whether there is or no a critical point in this energy range. The first round of NICA experiments will explore observables as particle yields and spectra, event-by-event fluctuations of multiplicity and transverse momentum as well as the corresponding joint distributions, all of which have already been employed in other experimental programs at RHIC and LHC and will offer a wealth of parameters in the higher density region.

\section{QCD in a Magnetic Field} \label{SII}
 
The experiments planned at NICA will allow exploring higher order susceptibilities and charge-distribution correlations in the region of high density and low temperatures.  As known, the fluctuations and correlations of conserved charges, as baryon number, electric charge, and strangeness, are sensitive to the degrees of freedom of the strongly interacting matter at finite temperature. They behave quite differently when QCD matter is in the hadronic or in the QGP phases \cite{Jeon}-\cite{Koch}.  Fluctuations and correlations are enhanced near the QCD phase transitions, and are related to the critical behavior of the QCD plasma \cite{Stephanov} - \cite{Skokov-1}. On the other hand, the fluctuations and correlations of conserved charges can be measured with event-by-event fluctuations in heavy ion collision experiments \cite{Koch}, \cite{Wang}, \cite{Abelev}, \cite{Abelev-1}, and hence are valuable probes of deconfinement and chiral-symmetry-restoring phase transitions \cite{Ejiri}-\cite{Ejiri-3}.

Fluctuations and correlations of conserved charges in the absence of  magnetic fields have been studied in the 2+1 flavor PNJL model \cite{Wei-jie}-\cite{Wei-jie-1} and the obtained results are consistent with those found in lattice simulations with an improved staggered fermion action with almost physical up and down quark masses and a physical value for the strange quark mass \cite{Cheng}.  
 
On the other hand, strong magnetic fields will likely be also generated in the off-central collisions at NICA, making it possible to explore the region of higher densities under a magnetic field. Even though the fields generated at lower energies will be weaker, since the typical field strength is $\sim\sqrt{s}$, they will last longer. The production of such magnetic fields in heavy-ion collisions and their potential for triggering new experimentally observable effects in QCD highlight the need for studying the properties of QCD and QCD-inspired theories in their presence. There are studies that indicate that the fluctuations and correlations of conserved charges are very sensitive to the external magnetic field. For example, in \cite{Fu} it was shown that the magnetic field increases the fluctuations and correlations in the regime of chiral crossover, making the transition of quadratic fluctuations more abrupt, and the peak structure of quartic fluctuations more pronounced. This in turn could affect the character of the transition and the location of the critical point \cite{CEP-B}- \cite{CEP-B-1}. Since the fluctuations and correlations of conserved charges can be observed in heavy ion collision experiments, it is natural to expect that the effects of strong magnetic fields produced in the early stage of off-central collisions will be imprinted onto these observables. 

Moreover, on the theoretical side, recent QCD-lattice calculations have found that the critical temperature for the chiral transition decreases with the magnetic field \cite{IMC-Lattice}-\cite{IMC-Lattice-2}, a phenomenon that has been termed inverse magnetic catalysis (IMC) and has motivated many theoretical works aimed to explain its physical origin \cite{IMC-Theory}-\cite{IMC-Theory-20}. It would be interesting to find a way to test the IMC lattice results on the experiments.

Other lines of efforts also indicate that the physics of dense QCD in a strong magnetic field is quite non-trivial and interesting. Among them we can mention the influence of the magnetic field on the QCD phases at very high density. At asymptotically large densities the most favored phase of QCD is the so-called Color-Flavor-Looked (CFL) superconducting phase \cite{CFL}. In this phase. a combination of the original electromagnetic potential and the eight gluon remains massless, thereby defining an in-medium, rotated electromagnetism \cite {CFL}, \cite{CFL-EM}. As a consequence, a conventional magnetic field can actually penetrate the color superconductor through its rotated component. With respect to the rotated electromagnetism, quarks have only integer charges and some of them are neutral; moreover, some of the gluons are also charged in this in-medium electromagnetism \cite{Cristina}, \cite{RevCS-B}. 

The implications of all the above is that an external magnetic field can significantly change the QCD phases at very high densities \cite{RevCS-B}. At weak magnetic fields, the color superconductor is approximately in the CFL phase because the charged Goldstone bosons that are part of the low-energy content of the CFL phase, are massive, but very light, so they cannot decay in pairs of quark and antiquark. When the field strength is of the order of the quarks' energy gap, the charged mesons become heavy enough to decouple and the low-energy physics is driven by five neutral massless bosons \cite{MagPhases}-\cite{MagPhases-1}. This low-energy corresponds to a different phase, the Magnetic-CFL phase \cite{CS-B}- \cite{CS-B-3}, with different transport properties.  Moreover, at fields comparable to the magnetic masses of the charged gluons, a chromomagnetic instability is developed that leads to the formation of a gluon-vortex state and to the antiscreening of the magnetic field \cite{Vortex}-\cite{Vortex-1}. The vortex state breaks the rotational symmetry in the plane perpendicular to the external magnetic field, hence the vortex formation corresponds to a phase transition from the Magnetic-CFL to the so-called Paramagnetic-CFL phase \cite{MagPhases}. The appearance of all these phases with increasing field illustrates the relevance of magnetic field effects on the properties of strongly interacting matter at finite densities.  A magnetic field is known to produce many other significant effects in color superconductivity \cite{CS-B1}-\cite{CS-B1-6} (for a review see \cite{RevCS-B}); it influences the BCS-BEC crossover \cite{BCS-BEC}-\cite{BCS-BEC-1}; it helps to test the topological properties of the QCD vacuum \cite{NPA803},\cite{PRD78}; and modifies the chiral inhomogeneous phases from intermediate to large densities \cite{KlimenkoPRD82}-\cite{Incera-1505}.

\section{Anomalous Transport in P-odd Quark Matter in a Magnetic Field} \label{SIII}

The interaction between quark matter and the large magnetic fields generated in off-central collisions opens a window to probe topological properties of the QCD ground state through a variety of anomalous transport phenomena \cite{anomaloustransport}-\cite{anomaloustransport-1}.  These phenomena include, in the case of hot quark matter, the Chiral Magnetic Effect (CME), \cite{NPA803},\cite{PRD78}, \cite{ChMEf}- \cite{ChMEf-2}; the Chiral Separation Effect (CSE) \cite{CSE}-\cite{CSE-1}; the Chiral Electric Separation Effect (CESE) \cite{CESE}; and the Chiral Magnetic Wave (CMW) \cite{CMW}. They are characterized by anomalous transport coefficients associated to the matter's response to externally applied electromagnetic fields. 

The most famous of the Parity-odd transport effects in hot QCD is the CME. In heavy-ion collisions, the CME can occur in the P-odd quark-gluon plasma that forms at high temperatures. The effect manifests as a separation of charge in the direction parallel to the magnetic field due to an anomalous current induced by the chiral imbalance of the medium and the lack of spin-degeneracy of the quarks in the lowest Landau level (LLL). The magnetic field then serves as a probe of the chiral quarks and hence of the gauge field fluctuations that are enhanced by sphaleron transitions at finite temperature. The CME is mainly relevant at temperatures large enough to allow over-the-barrier type of transitions (sphaleron-induced transitions) between topologically different QCD vacua. 

The CME effect can only be measured on an event-by-event basis because QCD does not violate parity globally, hence the "axial" chemical potential needed for it to happen fluctuates from one event to another. The charge separation produced by the CME can be measured with the help of charge-dependent azimuthal correlations \cite{PRC70}
\begin{eqnarray}
\gamma_{\alpha \beta}=\langle\cos(\phi_i+\phi_j-2\Psi_{RP})\rangle_{\alpha\beta}\\
 \delta_{\alpha\beta}=\langle\cos(\phi_i-\phi_j)\rangle_{\alpha\beta}
\end{eqnarray}
with $\alpha,\beta= \pm$ labeling the species, $\phi_{i,j}$ the azimuthal angles of two final state charged hadrons, and $\Psi_{RP}$  the reaction plane angle. STAR \cite{STAR09-10}-\cite{STAR09-10-1} and  PHENIX \cite{PHENIX-10-11}-\cite{PHENIX-10-11-1} collaborations at RHIC have reported measurements of the azimuthal correlations consistent with the expectation of the CME. More recently, the STAR Collaboration \cite{STAR14} reported the measurement of the same azimuthal correlation at different beam energies. The main conclusion of the new measurement is that the CME-induced signal is reduced at lower energy. For energy smaller than 20 GeV, the same-sign and opposite-sign correlations start to overlap with each other. This is consistent with the fact that at low temperature the QCD topological transition rate is strongly suppressed and thus the CME is attenuated. However, these observables also suffer from elliptic flow driven background contributions and cannot be entirely attributed to CME. Hence, a current outstanding challenge is to quantitatively decipher possible CME signals from the measured correlation observables. To address this issue, new measurements that would be sensitive to flow-driven versus magnetic-field-driven effects has been proposed \cite{1512.06602} to be combined with qualitatively computing both types of contributions for comparison with data.  The second phase of the beam energy scan (BES II) at RHIC and the experiments at NICA will provide additional data that should allow to finally settling this important question.

On the other hand, NICA will be able to explore the phase diagram of strongly interacting matter in the region of highly compressed baryonic matter and relatively low temperatures. Such a matter only exist in nature in the core of neutron stars. In the collisions planned at NICA, a large fraction of the beam energy will be converted into newly created hadrons and excitation of resonances whose properties may noticeably be modified by the surrounding dense medium. At very high densities, this hadron's mixture melts and its constituents, quarks and gluons, can form new phases of matter.  This high-baryon-density region is however out of the realm of lattice QCD, thus it has to be studied with model calculations. The model calculations have predicted a variety of dense quark matter phases, many of which are inhomogeneous in space and may have chiral symmetry broken in a different way than in the QCD vacuum. In this context, it is natural to ask ourselves if new anomalous transport phenomena can emerge in the phases that form in the dense and cold QGP and what new observables could be identified to allow us to probe the realization of these dense phases in the future experiments. 

This question will have to be approached in two steps. One will have to focus on the theoretical understanding of the possible phases and whether they possess or not anomalous transport properties. The second step will need to identify which observables are connected to the anomalous transport properties found in the step one and how these observables can be extracted from the experimental data. Both steps are important, but the second one will be needed to guide the experimentalists on what to look for and also to have quantitative predictions of the expected values of the parameters associated to a given observable. From the experience gained from the search for the CME in the hot QGP produced at RHIC and the issue of how to distinguish it from background effects, we expect that the tasks at high density will probably be at least as challenging and equally interesting. 
  
We have already identified an inhomogeneous phase of dense quark matter in a magnetic field that has been predicted by model calculations, is P-odd, and exhibits very interesting anomalous transport properties \cite{V-E-Topological}. It is characterized by a Dual Chiral Density Wave (DCDW) condensate \cite{DCDW} and requires the presence of a background magnetic field to generate the nontrivial topology that is behind the anomalous transport properties. As shown in \cite{V-E-Topological}, the interaction between the electromagnetic field and the DCDW matter is described by the equations of axion electrodynamics, a modification of the ordinary Maxwell equations proposed by Wilzcek many years ago \cite{axionElect}. The axion field in the DCDW medium depends on a spatial coordinate and is proportional to the modulation of the condensate. There might be other inhomogeneous phases of dense quark matter that could also exhibit axion electrodynamics \cite{Yamamoto}. 

The nontrivial topology of the DCDW phase arises because in the presence of the magnetic field, the spectrum of the lowest Landau level (LLL) fermions is asymmetric about zero \cite{PLB743}. This spectral asymmetry is measured by the Atiyah-Patodi-Singer topological invariant  $\eta_H=\lim_{s\to0}=\sum_k \mathrm{sgn}(E_k)|E_k|^{-s}$ \cite{AS}-\cite{AS-2} and gives rise to an anomalous contribution to the quark number density \cite{KlimenkoPRD82}-\cite{PLB743}.  As discussed in \cite{V-E-Topological}, once there is an anomalous quark number density, there will  also be an anomalous electric charge density $\rho^{elec}_A=\frac{e^2}{4\pi^2}qB$, and an anomalous Hall current $J_V= \frac{e^2}{2\pi^2} \nabla\theta\times \mathbf{E}$. The Hall current is perpendicular to both the direction of the applied magnetic field and the direction of the electric field that can be generated by the anomalous charge, even in the absence of ordinary (nonanomalous) electric charge. The anomalous Hall current, being perpendicular to $\mathbf{E}$,  is nondissipative. It will be important to carry out a detailed quantitative analysis of how the nondissipative Hall current could lead to observable signatures in off-central heavy-ion-collisions at higher baryon densities. These currents could deviate the quarks from the natural outward direction from the collision center and one would expect a different geometry of the particle flow in the DCDW phase in a magnetic field, compared to other possible dense quark matter phases with no anomalous Hall current. The realization of this P-odd phase in the QGP of the future NICA experiments is likely viable because the inhomogeneity of the phase is characterized by a length $\Delta x = \hbar / q \sim 0.6 fm$ for $q \sim \mu =300$ MeV \cite{KlimenkoPRD82}, much smaller than the characteristic scale $L\sim 10 fm$ of the QGP, while the time scale for this phase will be the same as for the QGP formed in the collision. To predict which observables can be connected to the anomalous Hall currents will require estimating the strength and direction of the anomalous Hall currents after the collision and figuring out if they can imprint any signature in measurable observables.

 If one takes into account the fluctuations of the axion field (equivalent to the fluctuations of the condensate modulation), the dispersions of the photon and the axion field become entangled.  For a linearly polarized photon with the electric field parallel to the background magnetic field, the coupling between the axion and the photon becomes linear. In such a medium light propagates as a linear combination of the axion and photon fields. Similar quasiparticle modes of coupled light/axion can be formed in topological insulators due to magnetic fluctuations \cite{axpolariton} and are known as polaritons. The special case of a polariton formed by coupled light and axion fields is called axionic polariton \cite{axpolariton}, so the fluctuations of the modulation parameter in the DCDW medium gives rise to an axion polariton. The dispersion of the axion polariton is gapped at certain frequencies, meaning that shining light within that range of frequencies in the medium leads to attenuated reflexion because it cannot propagate. The mass, stiffness, and velocity of the axion fluctuations that enter in the polariton's dispersion  depend on the properties of the DCDW medium and as such are functions of measurable quantities such as density, temperature, and magnetic field. The existence of the polariton modes opens the possibility for potentially observable signatures of the  DCDW phase that could be explored in off-central collisions at NICA experiments by shining laser light of various frequencies into the collision region and measuring the propagation of that light in this medium.

\section{Concluding Remarks } \label{SIV}

In summary, the availability of NICA at JINR will offer the opportunity to significantly enhance the current picture of the QCD phase map and improve our understanding of the effects of strong magnetic fields in dense quark matter.  These studies are directly relevant to the physics of heavy-ion collisions, but they also have a broader relevance, as they will be intertwined with and can mutually benefit from investigations in astrophysics and condensed matter systems. 

The importance of this topic is reflected in the increasing attraction the effects of an external magnetic field in quark matter is getting in the literature, e.g. in chiral magnetization \cite{ChMag}-\cite{ChMag-3}; modification of the nature of the chiral phase transition (e.g. from a crossover to a first-order phase transition) \cite{ChT}-\cite{ChT-6}; the chiral magnetic effect \cite{NPA803}, \cite{PRD78}, \cite{ChMEf}; spontaneous creation of axial currents \cite{CSE}; formation of $\pi_0$-domain walls \cite{DW}; modification of the location and nature of deconfinement \cite{B-decof}-\cite{B-decof-6}; formation of double quarkyonic chiral spirals \cite{APP5}, or inhomogeneous phases with chiral density waves in a magnetic field \cite{KlimenkoPRD82}, \cite{PLB743}, \cite{Incera-1505}, with interesting topological effects \cite{V-E-Topological}; just to name a few.

The planned experiments at NICA will allow us to use magnetic fields to probe the higher density region of the QCD map and use it to extract information about the topological content of the QCD phases in that region, explore new anomalous transport phenomena, and determine the phases that replace the quark-gluon plasma with increasing density. To tackle these problems will require coordinated collaborative efforts of theorists and experimentalists on which NICA no doubt will play a major role. 

\textbf{Acknowledgment}
\\
The authors acknowledge support from DOE Nuclear Theory  grant DE-SC0002179.

\end{document}